\documentclass[conference]{IEEEtran}
\IEEEoverridecommandlockouts

\usepackage{cite}
\usepackage{amsmath,amssymb,amsfonts}
\usepackage{algorithmic}
\usepackage{graphicx}
\usepackage{textcomp}
\usepackage{xcolor}
\usepackage{tikz}
\usepackage{float}
\usepackage{fancyhdr}

\usetikzlibrary{quantikz}
\def\BibTeX{{\rm B\kern-.05em{\sc i\kern-.025em b}\kern-.08em
    T\kern-.1667em\lower.7ex\hbox{E}\kern-.125emX}}
\fancypagestyle{specialfooter}{%
 \fancyhf{}  \fancyfoot[R]{ \noindent\fbox{%
 \parbox{\textwidth}{%
  {\footnotesize This work has been submitted to the IEEE for possible publication. Copyright may be transferred without notice, after which this version may no longer be accessible.} } }} }    
\begin{document}

\title{Benchmarking the Variational Quantum Eigensolver using different quantum hardware
\thanks{This research is part of the Munich Quantum Valley,
which is supported by the Bavarian state government with funds from the
Hightech Agenda Bayern Plus.}}

\author{
 \IEEEauthorblockN{Amine Bentellis\IEEEauthorrefmark{2}\IEEEauthorrefmark{1},
 Andrea Matic-Flierl\IEEEauthorrefmark{1}, Christian B. Mendl\IEEEauthorrefmark{2}\IEEEauthorrefmark{3}, Jeanette Miriam Lorenz\IEEEauthorrefmark{1}}
  \IEEEauthorblockA{\IEEEauthorrefmark{1}Fraunhofer Institute for Cognitive Systems IKS, Munich, Germany}
 \IEEEauthorblockA{\IEEEauthorrefmark{2}Technical University of Munich, CIT, Department of Computer Science, Garching, Germany}
\IEEEauthorblockA{\IEEEauthorrefmark{3}Technical University of Munich, Institute for Advanced Study, Garching, Germany}

 \{amine.bentellis, andrea.matic-flierl, jeanette.miriam.lorenz\}@iks.fraunhofer.de \\
 christian.mendl@tum.de

}

\maketitle

\thispagestyle{specialfooter}

\begin{abstract}
The Variational Quantum Eigensolver (VQE) is a promising quantum algorithm for applications in chemistry within the Noisy Intermediate-Scale Quantum (NISQ) era. The ability for a quantum computer to simulate electronic structures with high accuracy would have a profound impact on material and biochemical science with potential applications e.g., to the development of new drugs. However, considering the variety of quantum hardware architectures, it is still uncertain which hardware concept is most suited to execute the VQE for e.g., the simulation of molecules. Aspects to consider here are the required connectivity of the quantum circuit used, the size and the depth and thus the susceptibility to noise effects. Besides theoretical considerations, empirical studies using available quantum hardware may help to clarify the question of which hardware technology might be better suited for a certain given application and algorithm. Going one step into this direction, within this work, we present results using the VQE for the simulation of the hydrogen molecule, comparing superconducting and ion trap quantum computers. The experiments are carried out with a standardized setup of ansatz and optimizer, selected to reduce the number of required iterations. The findings are analyzed considering different quantum processor types, calibration data as well as the depth and gate counts of the circuits required for the different hardware concepts after transpilation.
\end{abstract}


\section{Introduction}
One of the primary applications being explored for quantum computers is quantum chemistry. It has the potential to simulate weakly and strongly correlated molecules and materials \cite{PhysRevX.8.011044 , Ma2020}, which fall under the category of simulation problems \cite{RevModPhys.92.015003 , Preskill2018quantumcomputingin}. Numerous algorithms have been suggested for this purpose. Some of them rely on fault-tolerant quantum hardware computation, such as the Quantum Phase Estimation (QPE) algorithm, which is currently not feasible. Thus, big parts of the research effort is put into Noisy Intermediate-Scale Quantum (NISQ) compatible algorithms, such as Variational Quantum Algorithms (VQA). The Variational Quantum Eigensolver (VQE) \cite{Peruzzo2014} is an exemplary VQA, which estimates the eigenvalues and low-lying eigenstates of a given Hamiltonian. Its application is not restricted to ground state energy evaluations but it can also be used to determine optimal molecule geometries \cite{PhysRevA.104.052402}. It also finds its uses outside of quantum chemistry, and can be employed in a variety of problems that can be formulated as a Hamiltonian \cite{mohanty2023analysis}.

Despite the vast amount of potential applications of the VQE, it is currently a question for which particular applications the VQE will show benefits in practice. A quantum advantage in general has only be shown in theory so far, such as e.g., in Grover's and Shor's algorithms, and in academic examples like in Google's Sycamore experiment \cite{Arute2019}. A practical quantum advantage, i.e., an advantage of quantum computation over classical computation in an industrially relevant problem has however not yet been demonstrated. To eventually reach such a practical quantum advantage, it has to be demonstrated for the complete computation workflow which includes both parts of quantum and classical computation. In particular, NISQ algorithms require the use of classical optimizers, which need to work in tandem with the quantum computation part. Noise presents an additional problem within the NISQ era and may imply that a result of a quantum computation is diluted, if the quantum circuits contained within the calculation are too wide and deep (i.e., require too many qubits and gates). Currently, techniques such as error mitigation and postselection are being utilized to counteract noise. It can also create barren plateaus, which act as obstacles in the loss landscape and hinder the optimization process of variational algorithms \cite{Wang2021}. Therefore, it is mandatory within the NISQ era to only consider shallower circuits.

Considering in this context different quantum hardware concepts, these may be better or less well suited to execute a given quantum circuit. However, which kind of quantum algorithm and noise level is acceptable for a given application in dependence on specific quantum hardware concepts is a largely unanswered question. We therefore argue that a systematic application-driven benchmarking procedure is required, that considers all computation steps - may it be classical or quantum computation, and their integration - as well as different quantum hardware technologies and noise levels.
The benchmarking procedure is consequently multi-layered, beginning with the definition of the problem, detailing the steps of quantum and classical computation, transpilation to quantum hardware and eventually extends to a specific quantum device on which the algorithm is executed. Additionally, the selection of the optimizer, the ansatz design and the problem mapping needs to be considered.

This complete benchmarking procedure being a complicated collection of individual steps, we focus in this study here on the step of selecting the quantum hardware suited for a specific application of the VQE. Therefore, the objective here is to evaluate and assess the solutions of a VQE implementation provided by both superconducting and ion trap hardware, considering both qualitative and quantitative aspects. The findings may help in understanding if a certain hardware concept is more suited than the other for a specific application and may give indications into which directions quantum algorithms, the interplay of quantum and classical computation and quantum hardware need to be developed.

In the particular study within this paper, we concentrate on simulating the H$_2$ molecule as a basic use case for comparing superconducting and ion trap quantum hardware. Choosing the H$_2$ molecule provides an example, where the ground state can exactly be calculated theoretically  and which has been extensively investigated in the scientific literature. Therefore, despite its relatively simple nature, it provides an adequate foundation for evaluating the quality of the solutions. Our work does not include an analysis of the scaling capabilities of the quantum hardware.

\section{Hardware}\label{hardware}

For this study we opted to choose two different quantum processors, one processor of Alpine Quantum Technologies (AQT) \cite{aqt} and one of IBM Q \cite{ibmq}. The main decisive factor in selecting these two were that they were easily available to us. This however also implies that for the selection of the IBM Q processor, we could not choose the most recent processor with improved error mitigation, as prohibitively long waiting times in the queue made extended tests by us impossible. Equally, in perspective, we will include further quantum hardware of different technologies into our studies, such as e.g. neutral atom systems, which had not been available to us pursuing the work presented in this paper.

\subsection{AQT trapped ion quantum computer}
The processor \textit{aqt\_marmot} hosted by AQT \cite{aqt} is based on trapped $^{40}$Ca$^+$ ions. The \textit{aqt\_marmot} system supports a register size of up to 16 qubits featuring all-to-all connectivity. The native gate set comprises single-qubit gates with arbitrary rotation angles and axis. The entangling operation is a two-qubit gate with arbitrary rotation angle around the x-axis that can be implemented between any qubit pair. The hardware supports all major software development kits, where we use IBM Qiskit \cite{Qiskit} to implement the presented measurements. Furthermore, we choose $[rx,rz,rxx]$ as the basis gate set in Qiskit, since it closely resembles the native gate set of the \textit{aqt\_marmot} system. The error rates of single-qubit gates ($rx$ error) are approximately $3\cdot10^{-4}$ on average, whereas the error rate of the two-qubit gate ($rxx$ error) is around $1\cdot10^{-2}$ on average. The typical gate times are $15\,\mu$s for single-qubit gates and $200\,\mu$s for two-qubit gates. T1 and T2 times are respectively $1.14 \pm 0.06$ seconds and $0.452 \pm 0.068$ seconds, making the coherence/gate time ratio $10^3$ \cite{aqt_coherence}. 
\subsection{IBM superconducting computer}
IBM is a primary supplier of superconducting devices for quantum computing \cite{ibmq}, and their machines can be readily operated via their cloud services and Qiskit \cite{Qiskit}. All runs are performed on the \textit{ibmq\_manila} hardware, with minimal time intervals between them, so that the calibration of the hardware is similar for each run. The specific backend is a Falcon processor Falcon r.11L version 1.1.4. Other available IBM Q processors feature improved coherence properties, but show similar readout (with the exception of \textit{ibm\_sherbrooke}), single and double-gate errors (Table \ref{tab:calibration_data} shows the calibration data at the time of the study). Average T1 and T2 across the whole chip are 169 $\mu$s and 76 $\mu$s.
\begin{table}[htpb]
    \centering
    \caption{\textit{ibmq\_manila} calibration data showing 1 and 2-qubits gate error rates as well as the gate time.}

    \resizebox{0.9\columnwidth}{!}{\begin{tabular}{|c|c|c|c|c|c|}
    \hline
        \textbf{Qubit}  & \textbf{ID error} & \textbf{$\sqrt{x}$ (sx) error} & \textbf{Pauli-X error} & \textbf{CNOT error} & \textbf{Gate time (ns)} \\ \hline
        0 & 2.057e-4 & 2.057e-4 & 2.057e-4 & \textbf{0\_1:} 5.621e-3 &\textbf{0\_1:} 277.3 \\ \hline
        1 & 2.236e-4 & 2.236e-4 & 2.236e-4 &\begin{tabular}{@{}c@{}}\textbf{1\_2:} 1.544e-2 \\\textbf{1\_0:} 5.62e-3\\ \end{tabular}   & \begin{tabular}{@{}c@{}}\textbf{1\_2:} 469.3 \\ \textbf{1\_0:} 312.8\\ \end{tabular} \\ \hline
        2 & 1.795e-3 & 1.795e-3 & 1.795e-3 & \begin{tabular}{@{}c@{}}\textbf{2\_3:} 8.233e-3 \\\textbf{2\_1:} 1.544e-2\\ \end{tabular}  &  \begin{tabular}{@{}c@{}}\textbf{2\_3:} 355.6 \\ \textbf{2\_1:} 504.8\\ \end{tabular} \\ \hline
        3 & 2.025e-4 & 2.025e-4 & 2.025e-4 & \begin{tabular}{@{}c@{}}\textbf{3\_4:} 5.0636e-3 \\ \textbf{3\_2:} 8.233e-3\\ \end{tabular} & \begin{tabular}{@{}c@{}}\textbf{3\_4:} 334.2 \\ \textbf{3\_2:} 391.1 \\ \end{tabular}\\ \hline
        4 & 3.341e-4 & 3.341e-4 & 3.341e-4 & \textbf{4\_3:} 5.063e-3 & \textbf{4\_3:} 298.7 \\ \hline
    \end{tabular}}
    \label{tab:calibration_data}
\end{table}
\section{Experiments and Results}\label{results}

\subsection{The VQE algorithm}
The Variational Quantum Eigensolver (VQE) is a popular quantum chemistry algorithm for near-term quantum computers \cite{Peruzzo2014 , TILLY20221}. The VQE leverages the Rayleigh-Ritz variational principle, whereby a quantum algorithm is trained to find the ground state of a particular molecule.
VQE is aimed at finding the energy $E_G$ of a Hamiltonian $H$,
\begin{equation}
    \begin{aligned}
    &H\ket{\psi_i} = E_i \ket{\psi_i}, \textrm{with } i = G, 1, 2, .... \\&E_G \leq E_1 \leq ..., \ \braket{\psi_i}{\psi_j} = \delta_{ij} 
    \end{aligned}
\end{equation}
Using the Born-Oppenheimer approximation \cite{https://doi.org/10.1002/andp.19273892002}, $H$ is represented in second quantization as:
\begin{equation} \label{secondQuant H}
    H= \sum_{pq} h_{pq} a_p^\dagger a_q + \frac{1}{2} \sum_{pqrs} h_{pqrs} a_p^\dagger a_q^\dagger a_r a_s ,
\end{equation}
with $a$ and $a^\dagger$ are respectively the fermionic annihilation and creation operators. Subsequently, this Hamiltonian is mapped using Jordan-Wigner transformation to the qubit space \cite{Jordan1928}. After mapping it can be expressed in terms of Pauli strings $\sigma^i$ as $H = \sum_i c_i \sigma^i$ with the coefficients $c_i \in \mathbb{R}$.  The cost function can then be formulated as the expectation value of $H$ over a trial state $\ket{\psi(\theta)} = U(\theta)\ket{\psi_0}$, where $U(\theta)$ is an ansatz and $\ket{\psi_0}$ is an initial state. The objective is to minimize the cost function, which is achieved by adjusting the parameters $\theta$ of the ansatz $U(\theta)$.
\begin{equation}
    C(\theta) = \bra{\psi(\theta)} H \ket{\psi(\theta)}
\end{equation}
Therefore, the cost function $C(\theta)$ is obtained from a linear combination of expectation values of $\sigma_i$.
The Rayleigh-Ritz variational principle states that the cost function $C(\theta)$ is both accurate and significant, where $C(\theta) > E_G$ and when $\ket{\psi(\theta)}$ represents the ground state $\ket{\psi_G}$ of $H$, equality holds true.

\subsection{Ansatz and Optimizer}
\begin{figure}[htpb]
\centering
\begin{quantikz} 
\centering
\lstick{$\ket{0}$} & \gate{R_y(\theta_1)} &  \ctrl{1} &\qw&\qw& \targ{}&\qw \\
\lstick{$\ket{0}$} & \gate{R_y(\theta_2)}& \targ{}  & \ctrl{1}&\qw&\qw&\qw\\
\lstick{$\ket{0}$} & \gate{R_y(\theta_3)}&\qw& \targ{}& \ctrl{1}&\qw &\qw\\
\lstick{$\ket{0}$} & \gate{R_y(\theta_4)}&\qw&\qw& \targ{} & \ctrl{-3}&\qw
\end{quantikz}
\caption{\label{fig:fig_hwery}   RY-CNOT ansatz using rotations around the Y axis}
\end{figure}
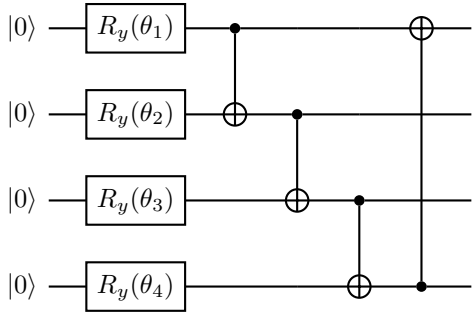

Two ansatzes were considered: the hardware-efficient ansatz (Figure \ref{fig:fig_hwery}) and the unitary coupled-cluster ansatz (UCC), which is chemically inspired. More specifically for the UCC ansatz, a variant based on single and double electron excitations (UCCSD) \cite{Peruzzo2014} was considered. While the UCCSD ansatz is a viable candidate, it suffers from several limitations that render it unsuitable for benchmarking purposes. Specifically, it exhibits poor scalability in terms of gate requirements, leading to increasingly deeper circuits as the molecule size increases. For the H$_2$ molecule the circuit is already relatively deep (circuit depth of 92 for \textit{ibmq\_manila}), and if moving to more complicated molecules like LiH, it becomes impossible to run on real hardware without simplifying the problem.
For the purposes of this work, we therefore employed the RY-CNOT ansatz. Figure \ref{fig:fig_hwery} illustrates that the circuit is composed of only four RY gates, which are parameterized, and an entangling layer that consists of CNOT gates arranged in a circular manner. This all-purpose ansatz not only simplifies the implementation on various quantum hardware with their unique gate sets but also results in a significantly faster optimization process.
 
\begin{figure}[htbp]
\centerline{\includegraphics[scale=0.3]{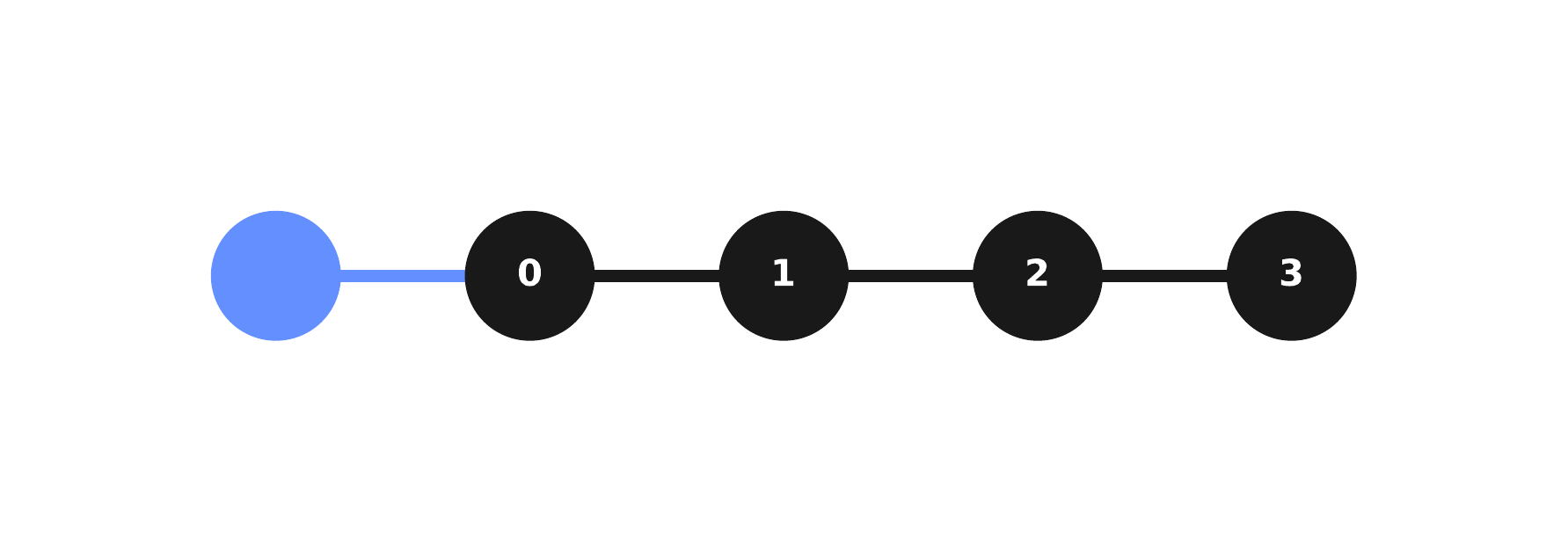}}
\caption{Chip topology for \textit{ibmq\_manila} comprising of five superconducting qubits.}
\label{fig:topo_manila}
\end{figure}

\begin{figure}[htbp]
\centerline{\includegraphics[scale=0.3]{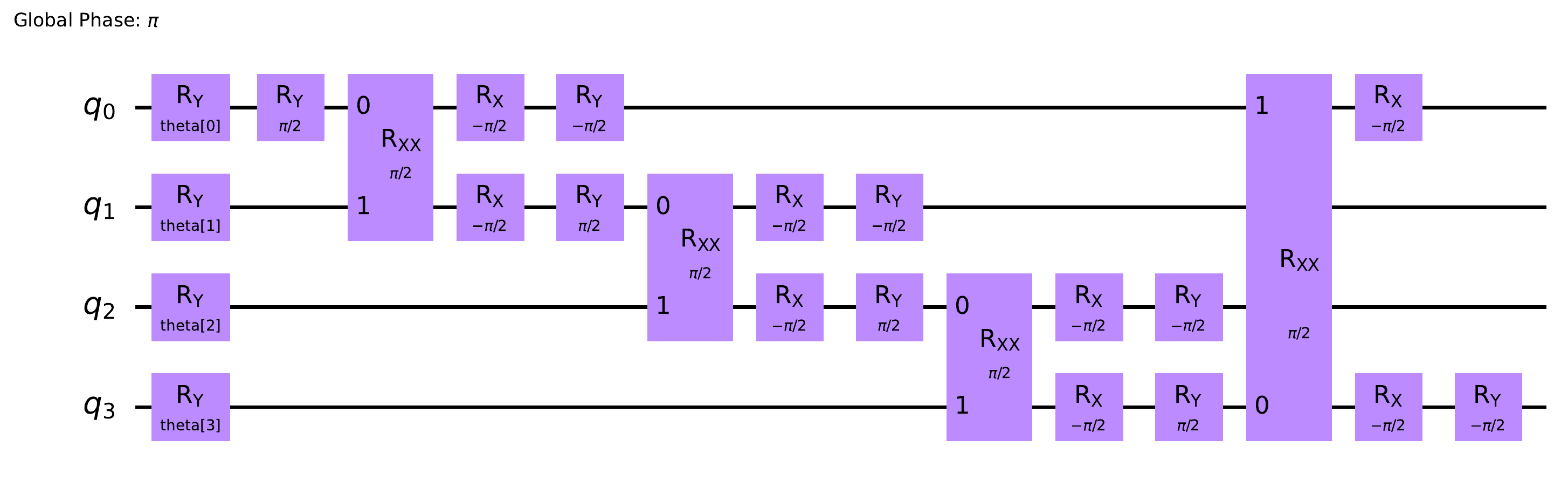}}
\caption{RY-CNOT circuit after transpilation on \textit{aqt\_marmot}.}
\label{fig:transpiled_AQT_circuit}
\end{figure}

\begin{figure}[htbp]
\centerline{\includegraphics[scale=0.255]{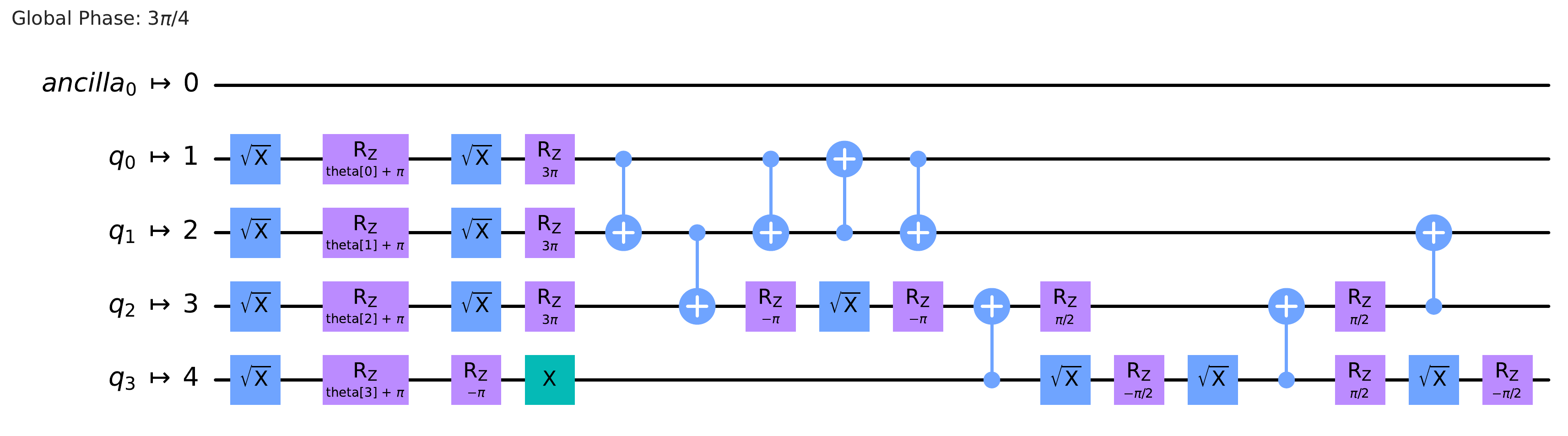}}
\caption{ RY-CNOT circuit after transpilation on \textit{ibmq\_manila}.}
\label{fig:transpiled_IBM_circuit}
\end{figure}

\begin{figure}[htbp]
\centerline{\includegraphics[scale=0.5]{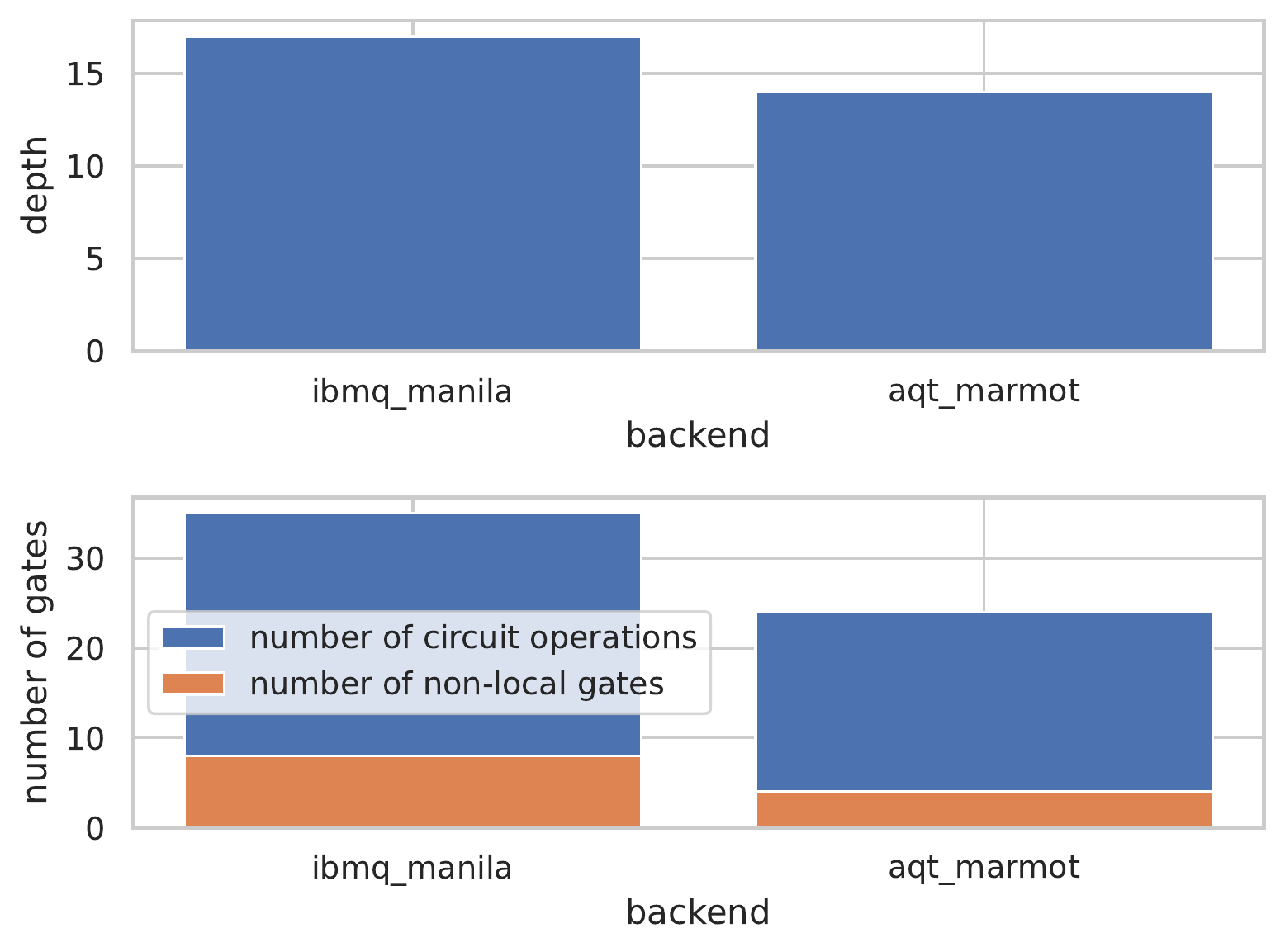}}
\caption{Depth and gate counts of post transpilation circuits.}
\label{fig:gate_count}
\end{figure}

After transpilation on \textit{aqt\_marmot} and \textit{ibmq\_manila} (resp. Figure \ref{fig:transpiled_AQT_circuit} and \ref{fig:transpiled_IBM_circuit}), both circuits significantly increase in size. For both circuits, Qiskit's optimization level 3 is used. Optimization level 3 spends the most computational effort to optimize the circuit out of the four available levels. Figure \ref{fig:gate_count} details the difference in terms of gate count and depth. We see circuits of similar depth (16 for \textit{ibmq\_manila} and 14 for \textit{aqt\_marmot}), however, the two transpiled circuits differ in the number of non-local gates or 2-qubit gates (8 for \textit{ibmq\_manila} and 4 for \textit{aqt\_marmot}). This discrepancy can be explained by the addition of two swapping operations for the IBM Q hardware since SWAP gates are, when transpiled, transformed into two consecutive and opposite CNOT gates. The use of SWAP gates is necessary depending on the architecture of a superconducting chip as these chips show a limited connectivity and can only directly entangle qubits with a direct connection on the chip (Figure \ref{fig:topo_manila} shows \textit{ibmq\_manila}'s topology).

The selection of an appropriate optimizer is a more nuanced decision that requires thorough testing. A comparison of several popular optimizers such as Nakanishi-Fuji-Todo (NFT) \cite{PhysRevResearch.2.043158}, Nelder-Mead \cite{10.1093/comjnl/7.4.308} and SPSA \cite{james_c_spall_implementation_1998} for VQE algorithms in this context is presented in the Appendix (Figure \ref{fig:opti}). All three optimizers evaluated avoid the need for computing the derivative of the circuit, which can be costly for large circuits, but instead rely on certain strategies to navigate the loss landscape. It is notable to say that gradient-based methods like Adam perform poorly on such tasks \cite{oliv2022evaluating}. Our results are consistent with \cite{oliv2022evaluating} and suggest that the NFT optimizer outperforms the other optimizers by a significant margin. Indeed, Figure \ref{fig:opti} suggests that NFT achieves convergence in an average of four iterations on this problem.

\subsection{Experiments}

For the simulations of the $H_2$ molecule we use the 4-qubit Hamiltonian:
\begin{equation*}
\centering
\begin{aligned}
    H &= c_1  I_0 \\ &+ c_2  Z_0Z_2 \\&+ c_3  Z_1Z_3 \\&+ c_4  (Z_3 + Z_1) \\&+ c_5 (Z_2 + Z_0) \\&+ c_6  (Z_2Z_3 +Z_0Z_1) \\&+ c_7 (Z_0Z_3 +  Z_1Z_2) \\&+ c_8  (Y_0Y_1Y_2Y_3 +X_0X_1Y_2Y_3 +Y_0Y_1X_2X_3 + X_0X_1X_2X_3)
\end{aligned}
\end{equation*}
using prefactors $c_i$ at a distance between atoms of 0.735 Å, where the specific values of $c_i$ can be found in the appendix. The Hamiltonian is represented using the minimal STO-3g basis set \cite{10.1063/1.1672392}. The exact ground state energy of the problem, calculated classically, is $E_{\text{FCI}} = -1.136189454088$ Ha. Despite the generic hardware-efficient ansatz utilized, a shot-based simulator (\textit{qasm\_simulator}) can achieve relatively accurate energies ($\lvert E_{\text{VQE}} - E_{\text{FCI}} \lvert = 0.00365$ Ha), indicating that the ansatz is capable of finding the solution. The experimental settings chosen for all the runs in this study are:
\begin{itemize}
    \item Circuit: RY-CNOT (c.f. Figure~\ref{fig:fig_hwery})
    \item Optimizer: NFT
    \item Framework: Qiskit
    \item Number of shots: 200
    \item No error mitigation
    \item Number of iterations: 15 
\end{itemize}
Error mitigation has emerged as a one of the key ways to deal with noise in NISQ era devices \cite{TILLY20221}. However, AQT's hardware currently does not support the native use of error mitigation techniques. Therefore, results for \textit{aqt\_marmot} are provided without error mitigation.     
    
\begin{table*}[htpb]
    \caption{Quantitative comparison of the two tested quantum processors}
    \centering
    \begin{tabular}{|l | l | l | l | l|}
    \hline
         & Final energy (Ha) & Minimum energy (Ha) & $\lvert E_{\text{VQE}} - E_{\text{FCI}} \lvert $ (Ha) & Time taken (quantum) (seconds)   \\ \hline
        \textit{ibmq\_manila} & $-0.975 \pm 0.032 $ & $-1.006 \pm 0.019 $   & $ 0.130 \pm 0.019 $ & $334 \pm 40 $ \\ \hline
        \textit{aqt\_marmot} & $-1.059 \pm 0.028 $ & $-1.100 \pm 0.010 $ & $ 0.036 \pm 0.010 $ & $13297 \pm 625 $ \\ 
        \hline
    \end{tabular}
    \label{tab:res_simulation}
\end{table*}

\begin{figure}[htbp]

\centerline{\includegraphics[scale=0.5]{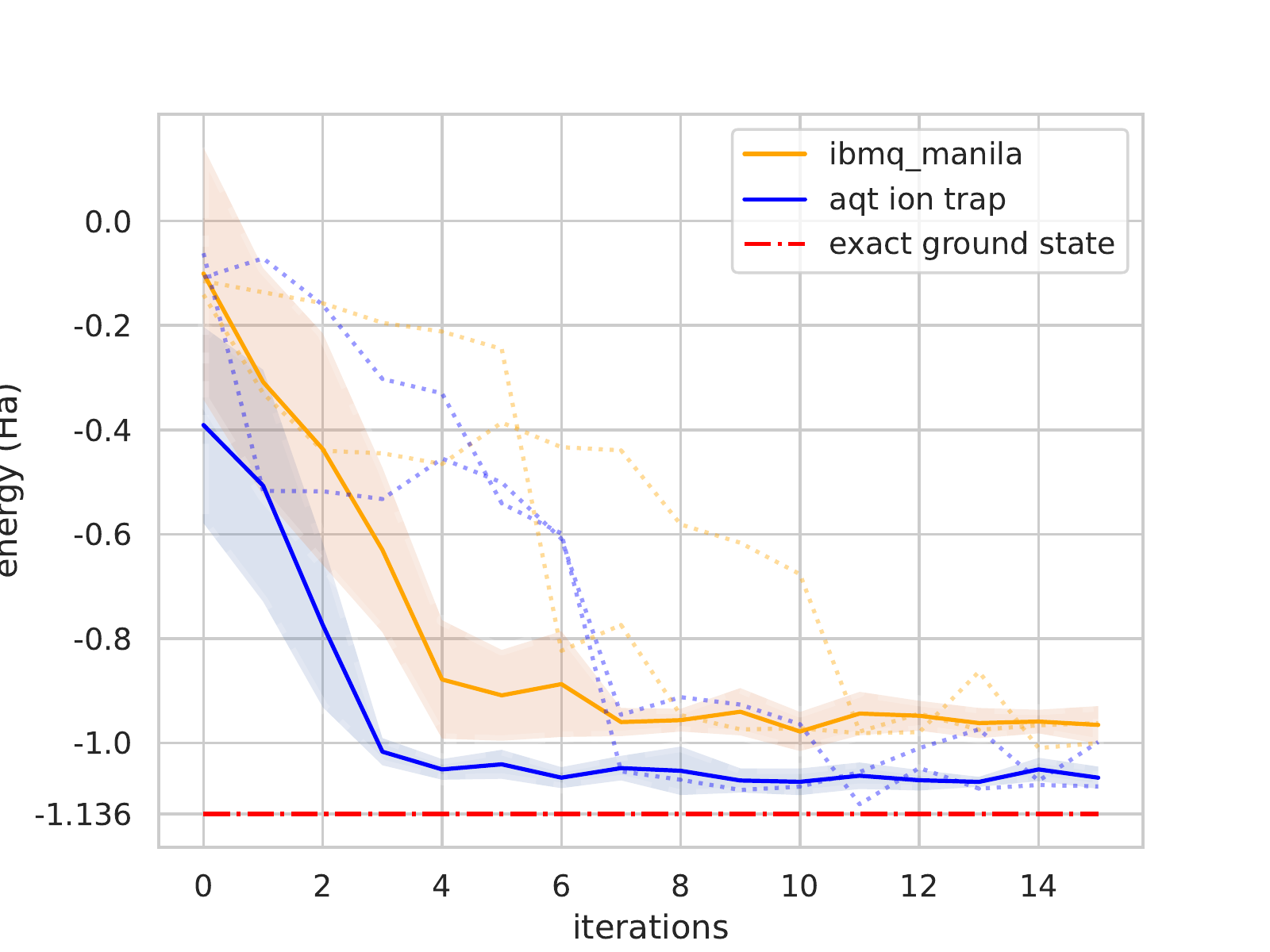}}
\caption{Comparison of the optimization process between the processors \textit{aqt\_marmot} and \textit{ibmq\_manila}.}
\label{fig:aqtvibm}
\end{figure}

This comparison is based upon nine different parameter seeds (i.e. parameters drawn from a uniform distribution of $[\pi, -\pi]$).
We see that the processor \textit{ibmq\_manila} takes an average of 7 iterations
to find the ground state, whereas the \textit{aqt\_marmot} requires approximately 4
iterations (Figure \ref{fig:aqtvibm}). In both cases, the choice of the seed can impact the optimization process, as seen with the two outliers (dotted lines in Figure \ref{fig:aqtvibm}). Furthermore, the effect of noise on the optimization process is demonstrated by the varying sizes of the uncertainty bands. These bands in Figure \ref{fig:aqtvibm}, which exclude the outliers, represent one standard deviation from the mean of the seven runs. Moreover, a mean energy difference of 0.094 between the two processors (or 9.34\%) is observed when considering all
runs (Figure \ref{fig:aqtvibm}). The precise reason
for this discrepancy in accuracy is not fully understood, as factors such as gate
fidelity and connectivity may not fully account for the observed difference in result precision. 

\begin{figure}[htbp]
\centerline{\includegraphics[scale=0.5]{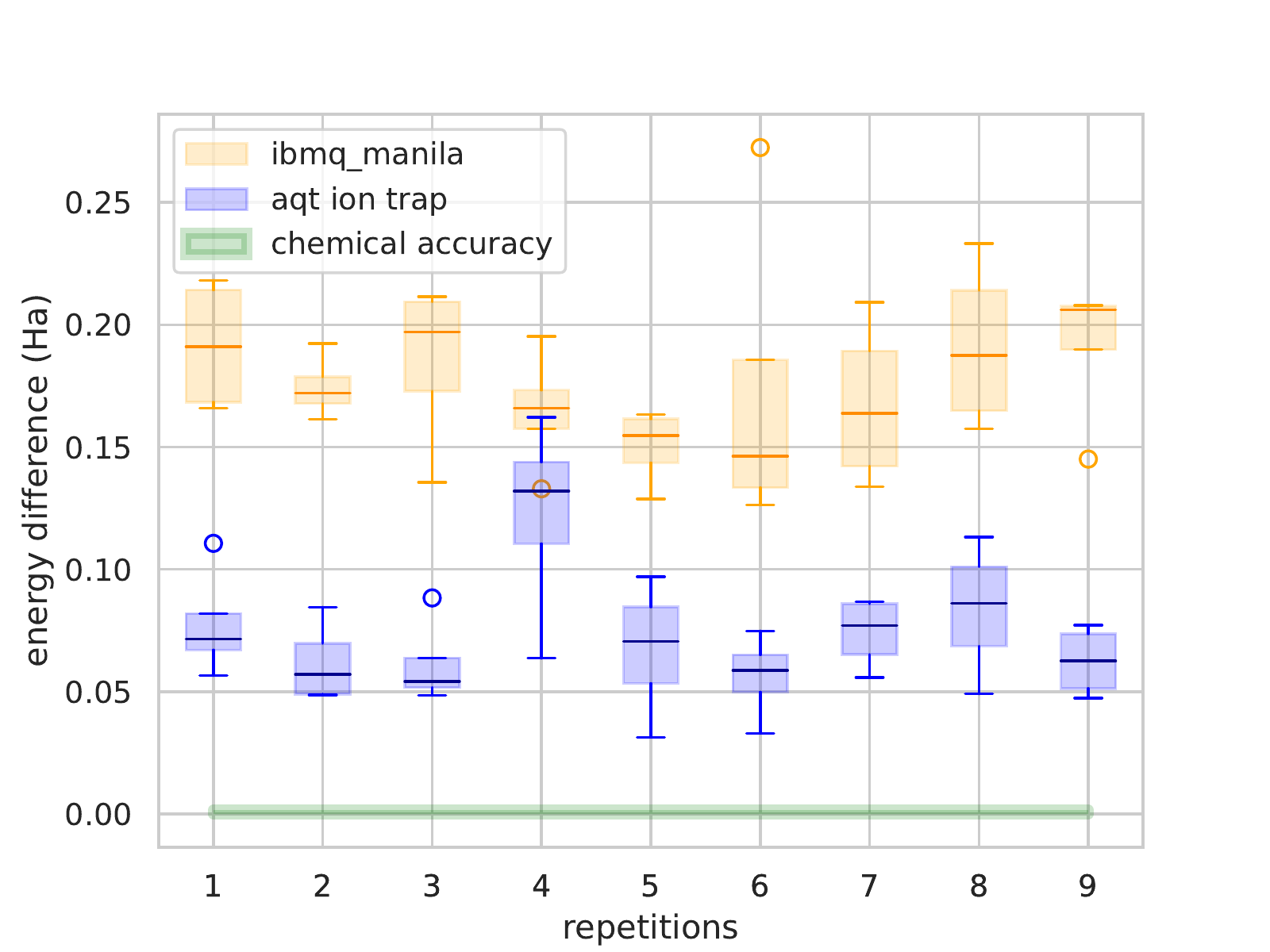}}
\caption{Energy difference between the theroretically known ground state and the last four iterations of the optimization process for all 9 runs (repetitions).}
\label{fig:energy_diff}
\end{figure}

Figure \ref{fig:energy_diff} and table \ref{tab:res_simulation} show the energy difference between the last steps of the optimization and the theoretically known ground state energy. This plot is generated from the last four iterations of respective hardware runs. This plot also shows that the unmitigated results are relatively close to the exact ground state.

One significant other parameter to take into consideration is the amount of time required for each hardware run. Over 15 iterations, one run on \textit{aqt\_marmot} takes on average 3 hours and 40 minutes, while \textit{ibmq\_manila} takes up
to 5 minutes to get the results (Table \ref{tab:res_simulation}). We only consider here the time required to evaluate the quantum circuit, and exclude the classical computation parts. It is however unclear how this runtime scales with increasing problem size.

\begin{figure}[htbp]
\centerline{\includegraphics[scale=0.5]{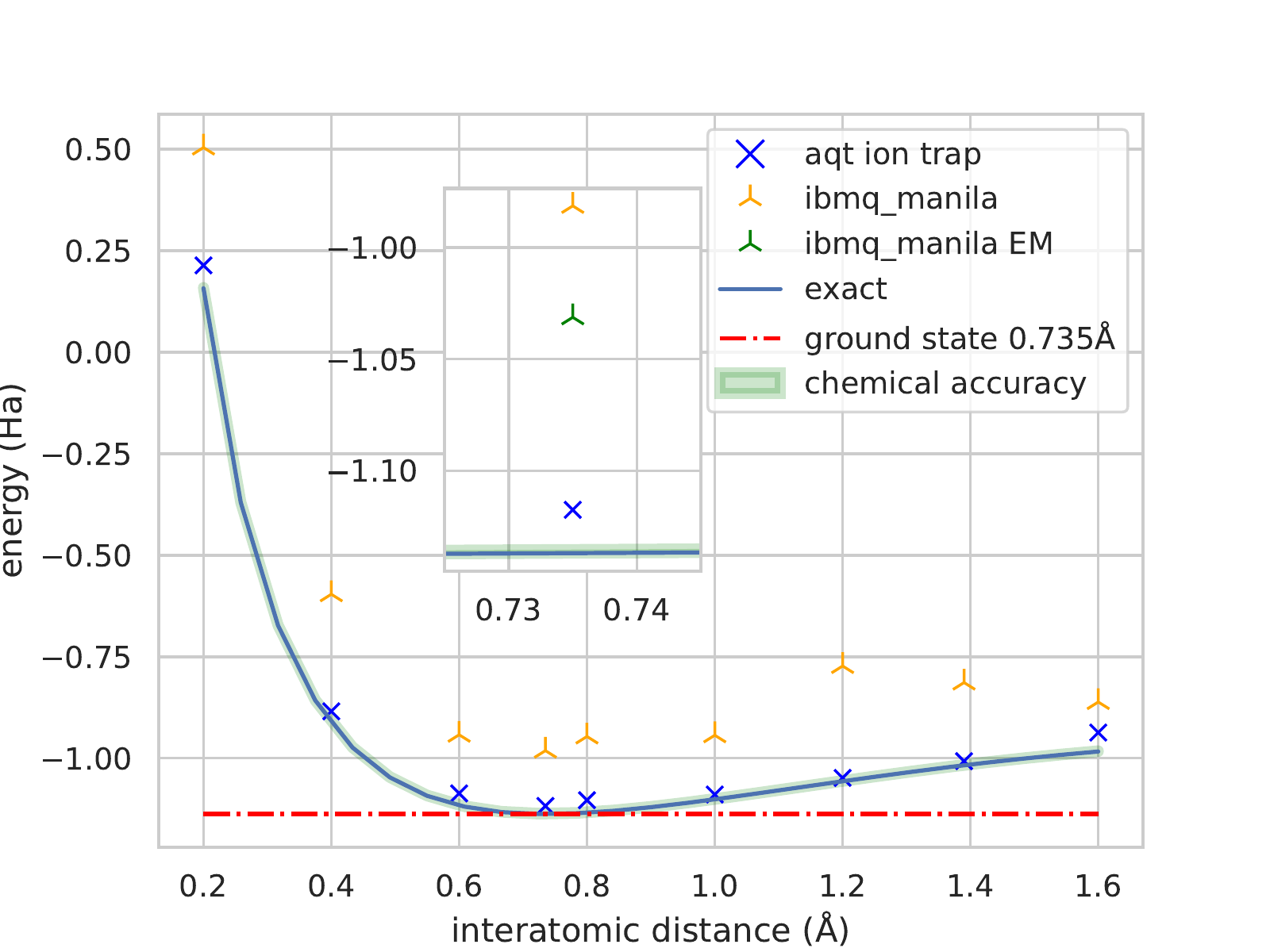}}
\caption{Ground state energy for different interatomic distances in Ångström. The minimum of the VQE optimization is taken for this plot.}
\label{fig:interatomic_diff}
\end{figure}

The internuclear distance plot shown in Figure \ref{fig:interatomic_diff} determines the molecular geometry, which serves as the foundation for many molecular property simulations.
The zoomed part of the plot corresponds to the global minimum located at 0.735 Å where \textit{ibmq\_manila} is compared to \textit{aqt\_marmot}. We also compare here the use of measurement error mitigation for the \textit{ibmq\_manila} hardware (green marker), which reveals an improvement in accuracy.

\section{Discussion} \label{discussion}

Quantum computation in the near future has restrictions regarding the overall number of operations that can be executed. Reducing circuit complexity is one of the motivations for benchmarking and application-based comparison, as we can identify the optimal software/hardware pairing. Comparing different quantum hardware processors, but also quantum algorithms, encompasses multiple interconnected aspects. The accuracy of quantum gate operations, the number of available qubits, the speed of primitive gates, and the duration of coherence times are all crucial fundamental measures for quantum computing hardware at a larger scale. Still, it remains challenging to predict how an algorithm will perform on a particular hardware concept based solely on hardware-close benchmarks.  

Here we have proposed the VQE in the context of quantum chemistry as a comparison basis. The ansatz used is problem-agnostic, so we can anticipate comparable performance for other minimization problems involving a sum of expectation values (such as Ising models). Our results demonstrate that all-to-all connectivity as present in ion trap hardware can significantly decrease the number of non-local gates required. We hypothesize that the optimized energy difference may be due to the impact of SWAP gates (which require 2-qubits gates) on the VQE's performance. Indeed, the cost of 2-qubit gates is much higher, both in terms of fidelity and execution time.

One other important aspect of benchmarking to consider is the time taken by the algorithm. When using variational algorithms we need to consider both the classical and the quantum execution times. The way the user interfaces with the different quantum hardware platforms can vary. The quantum systems of IBM Q can be accessed via cloud, which is not the case yet for \textit{aqt\_marmot}. Running algorithms through cloud services inherently introduces additional time overhead, but also makes the systems widely available. Besides the access time, the time that the quantum algorithm itself requires is more of interest to the study presented here. Here, a notable difference is observed between the two backends. Based on calibration data, it is known that \textit{ibmq\_manila} gate times are approximately $10^3$ times faster than \textit{aqt\_marmot}, which naturally results in a difference in quantum runtime. Another important factor to take into account is the coherence to gate time ratio. This metric not only quantifies the number of operations a quantum system can perform, but also reflects the number of gates that can be execute before the quantum state decays. The \textit{aqt\_marmot} backend, with its extended coherence time, can still execute meaningful operations despite having slower gate times.

It is worth noting that this study primarily focused on assessing the quality of the solution and did not delve into evaluating the scalability of the hardware. However, increasing the size of the molecule would directly increase the number of qubits required for the simulation, itself linearly increasing the number of SWAP operations required to connect the first and last qubit in superconducting hardware.

\section{Conclusion}\label{conclusion}

In conclusion, this study aimed to compare two specific quantum processors in the context of an example application from quantum chemistry using VQE. We have demonstrated that the VQE serves as a suitable benchmark for evaluating application-centered hardware performance. Specifically, we focused on assessing the solution quality achieved by two specific quantum processors, \textit{ibmq\_manila} and \textit{aqt\_marmot}. Results reported in this paper exclusivly apply to these two specific processor types and the specific quantum algorithm considered, and cannot directly be generalized to other quantum processors. Nevertheless, the findings contribute to the understanding of the capabilities and limitations of different quantum processors for executing variational algorithms. We hope that this study triggers further investigations into how different quantum hardware technology concepts perform for different quantum algorithms and application examples.

\section{Acknowledgements}
We would like to thank Alexander Erhard and Christian Sommer for their
help running the algorithms on the AQT processor \textit{aqt\_marmot} and providing the details of the backend. We acknowledge the use of IBM Quantum services for this
work. The views expressed are those of the authors, and do
not reflect the official policy or position of neither IBM nor AQT. The results of this work can also not be used as recommendation to use the one or the other quantum hardware technology, as the results are processor specific, calibration specific and depend on the precise quantum algorithm used, as well as depend on the precise classical optimizer.

\section{Appendix}

\subsection{Optimizers}
In the following, we briefly introduce the three optimizers being tested and a plot of the optimization process using the \textit{qasm\_simulator}. 
\begin{itemize}
    \item Nelder-Mead is a widely used gradient-free optimizer for classical optimization problems. It evaluates the loss function at the vertices of a simplex, updates the simplex based on the results, and continues iterating until it converges to the minimum of the loss function.
    \item The NFT algorithm optimizes parameters one at a time by utilizing the sine curve behavior of the loss function for each parameter. It requires two to three evaluations per iteration, based on a hyperparameter.
    \item Simultaneous Perturbation Stochastic Approximation (SPSA) is a gradient descent approximation.The gradient of a function is approximated by perturbing the parameters in a random way and using the resulting function evaluations to update the parameter estimates.
\end{itemize}
\begin{figure}[H]
\centerline{\includegraphics[scale=0.5]{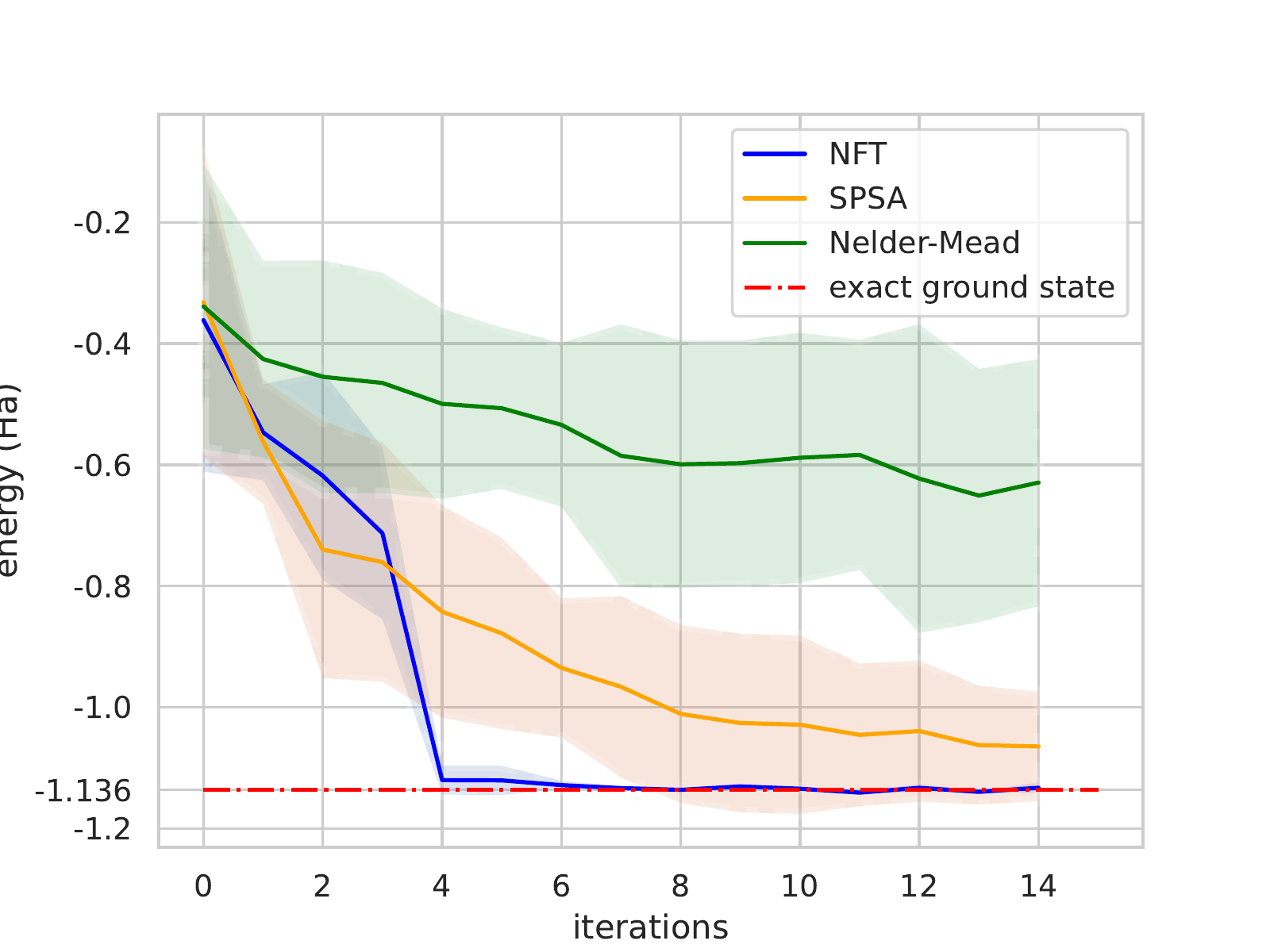}}
\caption{Popular optimizer comparison, the choice of these 3 optimizers is motivated by the work in \cite{oliv2022evaluating}. }
\label{fig:opti}
\end{figure}

\subsection{Hamiltonian Prefactors}
The Hamiltonian is generated using Qiskit's PySCF extension \cite{Qiskit}.
\begin{equation*}
\centering
\begin{aligned}
    c_1  &= -0.81054\\ c_2 &= 0.16614 \\c_3 &= 0.16892\\c_4  &= 0.17218\\c_5 &= -0.22573 \\ c_6 &= 0.12091\\c_7 &= 0.166145 \\ c_8 &= 0.04523
\end{aligned}
\end{equation*}
\bibliographystyle{unsrt}
\bibliography{main}
\end{document}